\documentclass[aps,pra,twocolumn,superscriptaddress,amssymb,showpacs]{revtex4}

\usepackage{graphicx}

\begin{document}


\title{Three-photon states in nonlinear crystal superlattices}



\author{D.~A.~Antonosyan}
\email[]{antonosyand@ysu.am}
\affiliation{Yerevan State University, Alex Manoogian  1, 0025,
Yerevan, Armenia}\affiliation{Institute for Physical Researches,
National Academy of Sciences,\\Ashtarak-2, 0203, Ashtarak,
Armenia}

\author{T.~V.~Gevorgyan}
\affiliation{Institute for Physical Researches, National Academy
of Sciences,\\Ashtarak-2, 0203, Ashtarak, Armenia}

\author{G.~Yu.~Kryuchkyan}
\email[]{kryuchkyan@ysu.am}
\affiliation{Yerevan State University, Alex Manoogian  1, 0025,
Yerevan, Armenia}\affiliation{Institute for Physical Researches,
National Academy of Sciences,\\Ashtarak-2, 0203, Ashtarak,
Armenia}


\begin{abstract}
It has been a longstanding goal in quantum optics to realize
controllable sources generating joint multiphoton states,
particularly, photon triplet with arbitrary spectral
characteristics. We demonstrate that such sources can be realized
via cascaded parametric down-conversion (PDC) in superlattice
structures of nonlinear and linear segments. We consider scheme
that involves two parametric processes:
$\omega_{0}\rightarrow\omega_{1}+\omega_{2}$,
$\omega_{2}\rightarrow\omega_{1}+\omega_{1}$ under pulsed pump and
investigate spontaneous creation of photon triplet as well as
generation of high-intensity mode in intracavity three-photon
splitting. We show preparation of Greenberger-Horne-Zeilinger
polarization entangled states in cascaded type-II and type-I PDC
in framework of consideration dual-grid structure that involves
two periodically-poled crystals. We demonstrate the method of
compensation of the dispersive effects in non-linear segments by
appropriately chosen linear dispersive segments of superlattice
for preparation heralded joint states of two polarized photons. In
the case of intracavity three-photon splitting, we concentrate on
investigation of photon-number distributions, third-order
photon-number correlation function as well as the Wigner
functions. These quantities are observed both for short
interaction time intervals and in over transient regime, when
dissipative effects are essential.
\end{abstract}

\pacs{42.65.Lm, 42.50.Dv, 42.65.Yj}

\maketitle

\section{INTRODUCTION}
Quantum communication and optical quantum computation rely on
controlled preparation of well-defined photonic states as well as
entangled states of electromagnetic modes. Generating of these
states is a primary task for the application of quantum
information processing. Particularly, the experimental
preparation, manipulation, and detection of multiphoton states are
of great interest for the implementation of quantum communication
schemes, quantum gates and for fundamental tests of quantum theory
\cite{Niels,Knill,GTit}. There exist a number of proposal for
implementing optical quantum-state engineering (see, for example,
reviewed paper \cite{Dell'}). Recently, this program has been
realized in details for single-photon and single-mode states. The
method of manipulating overall group delay mismatches between
interacting in a multilayered structures (superlattice structures)
by compensating dispersive effects in nonlinear segments by chosen
linear segments have been developed \cite{UrEd2,Kly,Gkl} for
synthesis of twin photon states. In this way, the demonstration of
heralded single photons prepared in pure quantum states from a
spontaneous parametric down-conversion (SPDC) has been done in
series of papers \cite{Uren} and engineering of quantum-optical
state has been demonstrated in \cite{Bim}. There are some attempts
to expand this strategy for producing heralded two-photon
polarization entangled states, since one promising approach to the
practical realization many of these tasks relies on qubits that
are encoded in the polarization states of single photons.  It was
shown that the production of one heralded polarization entangled
photon pair using only conventional down-conversion sources,
linear optical elements, and projective measurements requires at
least three initial pairs \cite{KokB}. An experimental scheme for
producing heralded two-photon entanglement, which relies on
triple-pair emission from a single down-conversion source was
proposed \cite{Silwa} and experimentally realized \cite{Barz}.

It has an important goal of quantum optics to realize effective
sources producing three-photon entangled states. Up to now,
several physical systems have been proposed for the direct
generation of photon triplet including cascaded spontaneous
parametric down-conversion \cite{10Green, 11Shih}, quantum dots
\cite{13Aichel} as well as electron-positron annihilation
\cite{Gupta}. Experimentally, three-photon down-conversion was
studied in third-order nonlinear media
\cite{Myun,16Douady,17Gravier} and also by using cascaded
second-order nonlinear parametric processes \cite{18Guo}. Most
recently, direct generation of photon triplets using cascaded
photon-pairs has been demonstrated in periodically poled lithium
niobate (PPNL) crystals \cite{Hubel}. This process will initiate
new class of experiments in quantum information technologies using
photons. Full characterization of a three-photon
Greenberger-Horne-Zeilinger (GHZ) state using quantum state
tomography has been performed \cite{Zeiling}. Very recently, the
distinction of three-photon GHZ and W states entangled in time and
space has been reported by comparing the second-order and
third-order correlation functions \cite{WRuub}. As an application
of this approach a  method of  generation of a narrowband
three-photon W state entangled in time (or energy) via two
four-wave mixing processes in cold atomic gas media has also been
proposed and studied \cite{JOSA}. It was also shown that
quantum-optical states based on three-photon down-conversion can
be effectively generated in cascaded optical parametric oscillator
(OPO). The cascaded OPO was proposed \cite{y,Zond} and
experimentally realized by using the dual-grid method of
quasi-phase matching (QPM) \cite{Kolk}. This scheme that involves
cascading second-order nonlinearities is based on parametric
processes of splitting and summing in which the frequencies
between the pump and two subharmonics frequencies are in the ratio
of 3: 2: 1. This cascaded configuration is different from the ones
proposed in the cited papers. It involves the fundamental mode
driven by an external pump field at the frequency $\omega_{0}$ and
two subharmonic modes $\omega_{2}=\frac{2\omega_{0}}{3}$, and
$\omega_{1}=\frac{\omega_{0}}{3}$. A remarkable feature of OPO is
comparatively low generation threshold \cite{y} in comparison with
the scheme of direct intracavity three-photon down-conversion,
where the pump power threshold is determined by third-order
susceptibility \cite{Myun}. Another remarkable feature is
formation of three phase locked states for both subharmonic modes
equally spaced by $\frac{2\pi}{3}$.

In this paper, we consider semiclassical- and quantum-properties
of photon triplets generated in cascaded parametric process noted
above: $\omega_{0}\rightarrow \omega_{1}+\omega_{2}$ and
$\omega_{2}\rightarrow \omega_{1}+\omega_{1}$. This setup offer a
variant of the scheme realized in the experiment \cite{Hubel} and
leads to generation of triple photons with approximately equal
frequencies. In this way two physical cases will be considered:
(a) cascaded spontaneous parametric three-photon splitting in a
"superlattice" of nonlinear and anisotropic linear segments pumped
by an ultrashort pulse; (b) amplification of these processes in
the presence of an optical cavity.

Thus, in the first part of this paper we propose detailed studies
of quasi-phase matching in cascaded configurations considering
complex nonlinear structures for simultaneously phase matching of
two parametric processes in optical lattices. We consider
dual-grid structure that involve periodically poled crystals for
effective generation of photon triplets as in the experiment
\cite{Hubel}, as well as we propose the analogical scheme for
preparation of the three-photon GHZ polarization entangled states.
We also consider dual-grid structures with second-order nonlinear
and linear materials for generation of photon triplet with
arbitrary joint spectrum. In this way we analyze the production of
heralded two-photon polarization entanglement in such superlattice
by using the conditional method of detection of auxiliary photons.
For this goal the various cases of the spectral factorization of
three-photon amplitude are analyzed in details.

The second part of the paper is devoted to studies of cascaded OPO
on the base of the above analysis. In the cited papers \cite{y,
Kryuchkyan2} the general theory of OPO has been developed,
however, without any concretization of the effective coupling
constants between modes subharmonics that are characterized by the
wave-vector mismatches of the parametric processes
$\omega_{0}\rightarrow \omega_{1}+\omega_{2}$ and
$\omega_{2}\rightarrow \omega_{1}+\omega_{1}$. Here, we consider
these points in details. Besides this, the lasing of
"three-photon" mode is investigated in transition through the
generation threshold. This investigation includes also analysis of
quantum statistics of modes i.e. photon number distributions,
third-order correlation functions and the Wigner function for
short interaction time as well as for over transition regime of
OPO, when dissipative effects are included.

The paper is arranged as follows. In Sec. II we derive the
Hamiltonian of the system and derive three-photon state through
QPM concurrent nonlinearities. In Sec. III we briefly describe
three-wave type-II interaction in nonlinear superlattice
structures, analyze generation of photon triplet in periodically
poled nonlinear crystals (PPNC) and propose the scheme for
preparation of GHZ-states. In Sec. IV we derive the conditions for
spectral factorization of the three-photon amplitude. Section V is
devoted to the analysis of cascaded OPO. We present the results on
photon-number distributions, third-order correlation function of
photon-numbers and the Wigner functions for subharmonic modes of
OPO for both short interaction time and steady-state regime.

\section{Three-Photon state through quasi-phase matching concurrent nonlinearities}

In this section, we consider multiple optical interactions in
$\chi^{\left(2\right)}$ nonlinear media, leading to the
simultaneous quasi-phase matching (QPM) of two parametric
processes. We investigate in details the cascaded processes of
three-photon splitting and summing that lead particularly to the
three-photon down-conversion. In the scheme the pump field at the
frequency $\omega_{0}$, which converts to the subharmonics at the
central frequencies: $\omega_{1}=\frac{\omega_{0}}{3}$ and
$\omega_{2}=\frac{2\omega_{0}}{3}$ throughout two cascaded
processes: $\omega_{0}\rightarrow \omega_{1}+\omega_{2}$ and
$\omega_{2}\rightarrow \omega_{1}+\omega_{1}$. Note, that a
complete description of photon triplet requires that all degrees
of freedom associated with the quantized electromagnetic field of
the occupied optical modes must be taken into account. The
specification of the photons involves their polarization, their
wave-vector $k$ and frequency or their spatial and time
dependence. The dispersion relationship between $k$ and $\omega$
reduces one degree of freedom, such that dependence from the
transverse wavevectors $k_{\bot}$ should be taken in account. In
this way, the transverse correlations in tripartite entanglement
have been studied in the paper \cite{Rubin}. Complete description
in which the source produces three-photon states in different
modes entangled in time and space has been done in the cited paper
\cite{WRuub}. Analogous multimode calculations for cascading
parametric systems in the layered media seem to be complicated and
can be carried out only by numerical means. However, the full
discussion of all degrees of freedom is beyond the scope of this
paper, for the further discussion we assume that the spatial modes
of the PDC photon pairs are independently decorrelated. We proceed
from the single-mode approximation that selects only a single
transverse spatial mode for the purpose of this paper. Thus, we
consider collinear, one-dimensional configurations focusing on
studies of spectral correlations and polarization entanglement.
Such spatial decorrelation can be achieved, for example, by
waveguides \cite{UrEd2}, in which the photons may be emitted only
into specific transversely-confined modes. On the other hand,
one-dimensional collinear approach used here is also fully
consistent for consideration of cascaded PDC, which generates
three-photon states in the presence of an optical cavity. This
analysis will be presented in Sec.V. Thus, in the discussed
approach the Hamiltonian of three-wave interaction reads as

\begin{eqnarray}
H(t)=S\sum_{\alpha,\beta,\gamma}\int^{L}_{0}dz\chi^{\left(2\right)}_{\alpha,\beta,\gamma}(z)E_{\alpha}(z,t)E_{\beta}(z,t)E_{\gamma}(z,t)=\nonumber\\
H^{\left(-\right)}(t)+H^{\left(+\right)}(t),~~~~~~~~~~\label{H1}
\end{eqnarray}
where $\chi^{\left(2\right)}_{\alpha,\beta,\gamma}(z)$ is the
second-order susceptibility of the one-dimensional medium of
length $L$. The electric field operator $E_{\alpha}(z,t)$ is the
superposition of the three components corresponding to the pump
field $E_{L,\alpha}(z,t)$ and the fields of two subharmonics
consisting of positive $E^{\left(+\right)}_{j,\alpha}(z,t)$ and
negative $E^{\left(-\right)}_{j,\alpha}(z,t)$ frequency electric
field operators ($j=1,2$). For simplicity, we also omit
polarization states of electromagnetic fields and polarization
indexes of nonlinear coefficient. The polarization states of modes
will be considered in sections III and IV. In this case, the field
operators obey
\begin{equation}
E^{\left(+\right)}_{j}(z,t)=i\int
\frac{d\omega}{2\pi}N_{j}(\omega)a_{j}(\omega)e^{i\left(k_{j}(\omega,z)z-\omega
t\right)},
\end{equation}
where $a_{j}(\omega)$, ($j=1,2$) are subharmonic modes operators,
while classical pump field reads as
\begin{equation}
E^{\left(+\right)}_{L}(z,t)=i\int
\frac{d\omega}{2\pi}E_{L}(\omega)e^{i\left(k_{L}(\omega,z)z-\omega
t\right)}.
\end{equation}
In these equations, $N_{j}(\omega)=\sqrt{\frac{\pi \hbar
\omega}{c\epsilon_{0}n^{2}_{j}(\omega)S}}$ is the normalized
factor, where $n_{j}(\omega)$ is the refractive index of the
medium at the given frequency, $S$ is the cross section area of
the beams, $k_{j}(z,\omega)=\frac{\omega_{j}}{c}n_{j}(z,\omega)$.
The fields have frequencies centered on carried frequencies
$\omega_{j}$ whose bandwidths $\Delta \omega_{j}$ are narrow in
comparison to the central frequencies $\omega_{0}$,
$\omega_{1}=\frac{\omega_{0}}{3}$ and
$\omega_{2}=\frac{2\omega_{0}}{3}$.

Note, that below we consider generation of photon triplet due to
cascading $\chi^{\left(2\right)}$ processes but not due to direct
$\chi^{\left(3\right)}$ interaction. In this approach the
Hamiltonian (\ref{H1}) is written trough the elementary
$\chi^{\left(2\right)}$ interactions that correspond to the
processes: $\omega_{0}\rightarrow \omega_{1}+\omega_{2}$ and
$\omega_{2}\rightarrow \omega_{1}+\omega_{1}$  while three-photon
down-conversion $\omega_{0}\rightarrow
\omega_{1}+\omega_{1}+\omega_{1}$ is calculated in the standard
manner (see, for example, \cite{WSAK, Hubel, JOSA}) in the
second-order of the perturbation theory on
$\chi^{\left(2\right)}$, (see, formula (\ref{PsD})). Thus, the
Hamiltonian of the system that describes nonlinear
$\chi^{\left(2\right)}$ interactions under the rotating-wave
approximation  and for neglecting the dependencies from the
transverse wave-vectors is calculated as
\begin{equation}
H(t)=H_{1}(t)+H_{2}(t),
\end{equation}

\begin{widetext}
\begin{eqnarray}
H_{1}(t)=-i\frac{S}{2\pi}\int d\omega d\omega_{1} d\omega_{2}
\zeta(\omega,\omega_{1},\omega_{2})N_{1}(\omega_{1})N_{2}(\omega_{2})e^{i\left(\omega_{1}+\omega_{2}-\omega\right)t}E_{L}(\omega)a^{+}(\omega_{1})b^{+}(\omega_{2})+h.c.,\nonumber\\
H_{2}(t)=-i\frac{S}{2\pi}\int d\omega^{'}_{2} d\omega^{'}_{1}
d\omega^{"}_{1}\xi(\omega^{'}_{1},\omega^{"}_{1},\omega^{'}_{2})N_{2}(\omega^{'}_{2})N_{1}(\omega^{'}_{1})N_{1}(\omega^{"}_{1})e^{i\left(\omega^{'}_{1}+\omega^{"}_{1}-\omega^{'}_{2}\right)t}a^{+}(\omega^{'}_{1})a^{+}(\omega^{"}_{1})b^{+}(\omega^{'}_{2})+h.c..\label{BHam}
\end{eqnarray}
\end{widetext}
Here and upwards, we use new denotations for the fields operators:
$a_{1}=a$, $a_{2}=b$. The Hamiltonian depends on the Fourier
spectra of the function $\chi^{\left(2\right)}(z)$
\begin{eqnarray}
\zeta(\omega,\omega_{1},\omega_{2})=\int^{L}_{0}dz\chi^{\left(2\right)}(z)e^{i\Delta
k_{1}(z)z},\label{zt}\\
\xi(\omega^{'}_{1},\omega^{"}_{1},\omega_{2})=\int^{L}_{0}dz\chi^{\left(2\right)}(z)e^{i\Delta
k_{2}(z)z},\label{qsi}
\end{eqnarray}
where the phase mismatch vectors are
\begin{eqnarray}
\Delta
k_{1}(z)=k_{L}(\omega_{0},z)-k_{1}(\omega_{1},z)-k_{2}(\omega_{2},z),\label{k1}\\
\Delta
k_{2}(z)=k_{2}(\omega_{2},z)-k_{1}(\omega^{'}_{1},z)-k_{1}(\omega^{"}_{1},z),\label{k2}
\end{eqnarray}
$\omega=\omega_{1}+\omega^{'}_{1}+\omega^{"}_{1}$, and $L$ is the
total length of the medium. This Hamiltonian is presented in the
general form that includes dispersion of interacting waves. Below
it will be concretized for multilayered nonlinear structures with
different susceptibilities of nonlinearity and with different
refractive indexes, pumped by the pulse laser light beam with
Gaussian distribution
\begin{equation}
E_{L}(\omega)=E_{0}\exp\Bigg(-\frac{\tau^{2}_{p}}{2}(\omega-\omega_{0})^{2}\Bigg).
\label{EG}
\end{equation}
We apply now Eq.(\ref{BHam}) for investigation of three-photon
states in the cascaded processes
$\omega_{0}\rightarrow\omega_{1}+\omega_{1}+\omega_{1}$. The
amplitude of this process can be calculated in the second-order of
the perturbation theory with Hamiltonian (\ref{BHam}). The vector
state in the second-order of the perturbation theory is described
in the Dyson series as
\begin{equation}
|\psi(t)\rangle=\left(\frac{-i}{\hbar}\right)^{2}\int^{t}_{-\infty}dt^{'}\int^{t^{'}}_{-\infty}dt^{"}H(t^{'})H(t^{"})|0\rangle_{a}|0\rangle_{b},\label{PsD}
\end{equation}
where $|0\rangle_{a}$ and $|0\rangle_{b}$ are vacuum states of the
modes at the frequencies $\omega_{1}$ and $\omega_{2}$. By using
the expressions (\ref{BHam}) we calculate the final state of this
cascaded setup for large time interval $t\rightarrow\infty$. By
using the commutation relations
$\left[b(\omega),b^{+}(\omega^{'})\right]=\left[a(\omega),a^{+}(\omega^{'})\right]=\delta(\omega-\omega^{'})$,
the result is calculated as
\begin{widetext}
\begin{eqnarray}
|\psi\rangle=\frac{S^{2}}{2\hbar^{2}}\int
d\omega_{1}d\omega^{'}_{1}d\omega^{"}_{1}N_{1}(\omega_{1})N_{1}(\omega^{'}_{1})N_{1}(\omega^{"}_{1})
\Phi(\omega_{1},\omega^{'}_{1},\omega^{"}_{1})|0_{b},1(\omega),1(\omega^{'})1(\omega^{"})\rangle,\label{Ps3}
\end{eqnarray}
\end{widetext}
where $|1(\omega)\rangle=a(\omega)|0\rangle$ is a single-mode
state, $0_{b}, 1(\omega)$ are the occupation numbers of the modes
and
\begin{widetext}
\begin{equation}
\Phi(\omega_{1},\omega^{'}_{1},\omega^{"}_{1})=\frac{i}{\pi}E_{L}(\omega_{1}+\omega^{'}_{1}+\omega^{"}_{1})\int
\frac{N^{2}_{2}(\omega_{2})\zeta(\omega,\omega_{1},\omega_{2})\xi(\omega^{'}_{1},\omega^{"}_{1},\omega_{2})}{\omega^{'}_{1}+\omega^{"}_{1}-\omega_{2}+i\varepsilon}d\omega_{2},\label{FiM}
\end{equation}
\end{widetext}
is the amplitude of three-photons at the frequencies
$\omega_{1},\omega^{'}_{1},\omega^{"}_{1}$,
($\varepsilon\rightarrow 0$). Three-photon down-conversion is
controlled by energy conservation between the pump and daughter
photons $\omega_{0}=\omega_{1}+\omega^{'}_{1}+\omega^{"}_{1}$,
while the amplitude is determined by the pump spectral amplitude
and the phase-matching functions.

For the case of degenerate three-photon down-conversion:
$\omega_{1}\simeq\omega^{'}_{1}\simeq\omega^{"}_{1}\simeq\frac{\omega_{0}}{3}$
the maximum of this amplitude is realized in the range of
resonance
$\omega_{2}=\omega^{'}_{1}+\omega^{"}_{1}=\frac{2\omega_{0}}{3}$.
In this case the Eq.(\ref{FiM}) approximately reduces to the
result
\begin{widetext}
\begin{equation}
\Phi(\omega_{1},\omega^{'}_{1},\omega^{"}_{1})=E_{L}(\omega)
N^{2}_{2}\left(\omega_{2}\right)\zeta(\omega,\omega_{1},\omega^{'}_{1}+\omega^{"}_{1})\xi(\omega^{'}_{1},\omega^{"}_{1},\omega^{'}_{1}+\omega^{"}_{1}).\label{FipD}
\end{equation}
\end{widetext}
Note, that the approximated result (\ref{FipD}) can be directly
obtained from Eq.(\ref{PsD}) if we use the Teylor type
time-ordering instead of the Dyson series expansion. The basis of
this approximation for the parametric processes has been also
discussed in \cite{Ralph,LMNR}. Below, we represent application of
these results for effective generation of photon triplet
\cite{Hubel}, for preparation of both three-photon GHZ entangled
states and the heralded polarization states.

\section{Three-photon splitting in composite nonlinear media}

In this section we first briefly discuss three-wave type-II
interaction in one-dimensional $\chi^{\left(2\right)}$ nonlinear
composite media, consisting of alternating layers with different
coefficients of nonlinearity. In this case, the interaction
Hamiltonian (\ref{H1}) can be expressed as the sum of interactions
in each layer in terms of the electric fields in $n$th layer,
$E_{Ln}(z,t)$, $E^{\left(-\right)}_{jn}(z,t)$

\begin{eqnarray}
H(t)=S\sum_{n}{\int^{z_{n+1}}_{z_{n}}{dz\chi^{\left(2\right)}(z)E^{*}_{pn}(z,t)E^{\left(-\right)}_{1n}(z,t)E^{\left(-\right)}_{2n}(z,t)}}\nonumber\\
+h.c.,~~~~\label{HGen}
\end{eqnarray}
where $\chi^{\left(2\right)}(z)$ is the second-order nonlinearity.
In approximation that the effects of refraction from the each
layers are small the classical pump field $E_{pn}$ and the
positive frequency parts of the fields of the subharmonics:
$E^{\left(-\right)}_{jn}(z,t)$ ($j=1,2$) can be expressed as
\begin{eqnarray}
E_{pn}(z,t)=\int d\omega e^{i\omega
t}E_{Ln}(\omega)e^{\left(-ik_{0n}(\omega)(z-z_{n})\right)},\label{Ep}
\end{eqnarray}

\begin{eqnarray}
E^{\left(-\right)}_{jn}(z,t)=\int d\omega e^{i\omega t}
A^{*}_{jn}(\omega)e^{-ik_{n}(\omega)(z-z_{n})}a^{+}_{j}(\omega).\label{En}
\end{eqnarray}

The electric fields (\ref{Ep}), (\ref{En}) have been studied in
\cite{WSAK} for one dimensional composite materials. In the case
of composite layered structure the coupling function: $\xi$
(and/or $\zeta$), (see Eqs.(\ref{zt}),(\ref{qsi})) could be
rewritten in the following form (see, the paper \cite{Kly})
\begin{eqnarray}
\xi=\sum_{n}\int^{z_{n+1}}_{z_{n}}\chi_{n}e^{i\Delta
k_{n}z}dz=\nonumber\\
\sum_{m}{l_{m}\chi_{m}e^{-i\left(\varphi_{m}+\frac{\Delta{k_{m}}l_{m}}{2}\right)}sinc
\left(\frac{\Delta{k_{m}}l_{m}}{2}\right)},\nonumber\\
\varphi_{m}=\sum^{m-1}_{n}{l_{n}\Delta k_{n}},~~~~~
\varphi_{1}=0.\label{klishkoEq}
\end{eqnarray}
Here, $\chi_{m}$ is the second-order susceptibility in the $m$-th
layer, $l_{m}$ is the length and $\Delta k_{m}$ is the phase
mismatch vector for the $m$-th layer.

It is easy to generalize these results on the case when
polarization structure of three-wave interaction is important.
Note, that in this case we should not include a summing on the
polarization indexes in the formula (\ref{H1}), if the phase
matching of three-waves interaction should take place in each
$m$th layer only for the concrete combinations of three indexes:
$\alpha_{m}$, $\beta_{m}$, $\gamma_{m}$. In this case, the tensor
of second-order susceptibility
$\chi^{\left(2\right)}_{\alpha,\beta,\gamma}$ reduced to
$\chi_{m}=\chi^{\left(2\right)}_{\alpha,\beta,\gamma}$
susceptibility in the $m$th layer. However, in this case the phase
mismatch vector for the $m$th layer $\Delta k_{m}$ also depends on
the indexes $\alpha_{m}$, $\beta_{m}$, $\gamma_{m}$. Note, that
this result is convenient for realizing various schemes of QPM, in
particular, for description of periodically poled nonlinear
structures. The technique of quasi-phase-matching (QPM) and the
availability of multiple gratings in a single periodically poled
nonlinear crystal (PPNC) offer flexibility in achieving multiple
interactions, which provide enhanced functionality of nonlinear
optical devices.

As has been mentioned in Sec. II the maxima of the three-photon
process depend on maxima of coupling constants of the each
cascading process. Perfect QPM is the condition for the effective
process implementation. Thus, in the subsection A QPM of cascading
three-photon process: $\omega_{0} \rightarrow
{\omega_{1}+\omega_{2}}$, $\omega_{2} \rightarrow
{\omega_{1}+\omega_{1}}$ is considered in the complex nonlinear
structures with dual-gratings, while the subsection B is devoted
to investigation of frequency-uncorrelated photon triplet and
subsection generation, C involves patterns of preparation
polarization entangled GHZ states in such structures.

\subsection{QPM for generation photon triplets}

In this subsection we develop the theoretical approach of the
cascaded generation of photon triplets, particularly, obtained
experimentally in \cite{Hubel}. We analyzed so called dual-grid
nonlinear crystal in frame of mismatching of two SPDC processes.
The crystal consists of two periodically poled nonlinear segments
with different periods. First segment involves $M$ domains of the
length $l_{1}$, with positive $\chi$ and negative $-\chi$
susceptibilities which are interchanged one to other and
refractive index $n_{1}$ (corresponding phase-matching function is
$\Delta k_{1}$), and the second segment consists of $N$ domains of
the length $l_{2}$ again with positive $\chi$ and negative $-\chi$
susceptibilities and refractive index $n_{2}$ (corresponding
phase-matching function is $\Delta k_{2}$). The crystal is pumped
by a laser field, thus the first segment become primary
down-conversion source, and photon pair is created. One of the
photons from this pair drives the secondary down-conversion
process in the second segment, generating a second pair, hence, a
photon triplet. Because the photon triplet originates from a
single pump photon, the created photons are strongly correlated.
The $\omega_{0} \rightarrow {\omega_{1}+\omega_{2}}$ and
$\omega_{2} \rightarrow {\omega_{1}+\omega_{1}}$ processes has
been implemented in the first and the second sections of the
crystal, correspondingly. We calculate the coupling constants of
parametric processes: $\zeta$ and $\xi$ as a function of the
finite number of nonlinear segments. Below, we consider spectral
amplitudes of these cascaded processes assuming the frequencies
$\omega_{1}$ and $\omega_{2}$ as variable quantities distributed
around the central values $\omega_{1}=\frac{\omega_{0}}{3}$ and
$\omega_{2}=\frac{2\omega_{0}}{3}$. The three-photon amplitude
$\Phi(\omega_{1},\omega^{'}_{1},\omega^{"}_{1})$ from
Eq.(\ref{FipD}) is proportional to the product of the coupling
constants and the pump field amplitude:
$E_{L}(\omega_{1}+\omega^{'}_{1}+\omega^{"}_{1})$, which is taken
in the form of Gaussian distribution (\ref{EG}). This lattice is
arranged in the manner that each of the Fourier functions
(\ref{zt}), (\ref{qsi}) is determines by one of the family of the
segments. The details of calculations based on formula
(\ref{klishkoEq}) can be found in \cite{WSAK}. We derive the
following expression for $\zeta$ in terms of Eq.(\ref{klishkoEq})
\begin{eqnarray}
\zeta=l_{1}\chi e^{-i\frac{L_{1}\Delta
k_{1}}{2}}sinc\left(\frac{l_{1}}{2}\Delta{k_{1}}\right)\frac{sin{\frac{Ml_{1}}{2}(\Delta
k_{1}-q_{1})}}{sin{\frac{l_{1}}{2}(\Delta
k_{1}-q_{1})}}+\nonumber\\
l_{2}\chi e^{-i\frac{L_{2}\Delta
k_{1}}{2}}sinc\left(\frac{l_{2}}{2}\Delta{k_{1}}\right)\frac{sin{\frac{Nl_{2}}{2}(\Delta
k_{1}-q_{2})}}{sin{\frac{l_{2}}{2}(\Delta k_{1}-q_{2})}},
\label{SumZt-Hum}
\end{eqnarray}
where
\begin{eqnarray}
\Delta k_{1}=
k_{L}(\omega_{0})-k_{1}(\omega_{1})-k_{2}(\omega_{2}).
\end{eqnarray}
Here $q_{1}=\frac{2\pi}{d_{1}}$, $q_{2}=\frac{2\pi}{d_{2}}$ are
the harmonic grating wave vectors for the first and the second
segments, respectively, $d_{1}=2l_{1}$, $d_{2}=2l_{2}$,
$L_{1}=Ml_{1}$ and $L_{2}=Nl_{2}$ are the total lengths of first
and second sections, correspondingly.

We assume that effective $\omega_{0} \rightarrow
{\omega_{1}+\omega_{2}}$ process goes in the first section, so the
conditions of the perfect QPM is $\Delta k_{1}=q_{1}$.
Correspondingly, it is easy to see that the second term in
(\ref{SumZt-Hum}) could be neglected proceeding from the QPM
condition we are required.

Let us write the approximation of (\ref{SumZt-Hum}) for $M\gg 1$,
$L_{1}\gg l_{1}$ and $\Delta k_{1}\simeq q_{1}$. We represent
phase-matching function in following form $\Delta k_{1}=\Delta
k^{\left(0\right)}_{1}+\delta k_{1}$, where $\Delta
k^{\left(0\right)}_{1}$ is the 0-order term of the Taylor series
and $\delta k_{1}$ is the function of the high-order terms. We
rewrite QPM condition in the following form $\Delta
k^{\left(0\right)}_{1}=q_{1}$, here is asseverated that the
constant term could be compensated by the grating wave vector
$q_{1}$. Using this approach and assuming that $l_{1}\delta k_{1}$
is very small and could be neglected we could find that the $sinc$
function in the (\ref{SumZt-Hum}) becomes a constant multiplier
equal to $\frac{2}{\pi}$. Thus, the final approximated form of the
$\zeta$ is represented in following form
\begin{equation}
\zeta=\frac{2}{\pi}Ml_{1}\chi e^{-i\frac{L_{1}\Delta
k_{1}}{2}}sinc\left(\frac{L_{1}}{2}(\Delta
k_{1}-q_{1})\right).\label{ApZt-Hum}
\end{equation}
This result is transformed to an expression that usually used for
PPNC in the standard treatment.

The Fourier function (\ref{qsi}) is calculated in the following
form
\begin{eqnarray}
\xi=l_{1}\chi e^{-i\frac{L_{1}\Delta
k_{2}}{2}}sinc\left(\frac{l_{1}}{2}\Delta{k_{2}}\right)\frac{sin{\frac{Ml_{1}}{2}(\Delta
k_{2}-q_{1})}}{sin{\frac{l_{1}}{2}(\Delta
k_{2}-q_{1})}}+\nonumber\\
l_{2}\chi e^{-i\frac{L_{2}\Delta
k_{2}}{2}}sinc\left(\frac{l_{2}}{2}\Delta{k_{2}}\right)\frac{sin{\frac{Nl_{2}}{2}(\Delta
k_{2}-q_{2})}}{sin{\frac{l_{2}}{2}(\Delta k_{2}-q_{2})}},
\label{Xi-Hum}
\end{eqnarray}
where
\begin{eqnarray}
\Delta
k_{2}=k_{2}(\omega_{2})-k_{1}(\omega^{'}_{1})-k_{1}(\omega^{"}_{1}).
\end{eqnarray}
Thus, we assume that effective $\omega_{2} \rightarrow
{\omega^{'}_{1}+\omega^{"}_{1}}$ process goes in the second
section, so the conditions of the perfect QPM is $\Delta
k_{2}=q_{2}$. It is easy to see that in this case the first term
in (\ref{Xi-Hum}) could be neglected proceeding from the QPM
condition we required. We could write the approximated form for
$\xi$ similar to Eq.(\ref{ApZt-Hum}).

So we calculated the coupling constants for the both SPDC
processes and found the QPM conditions for their perfect
implementation. The both conditions depend on grating wave
vectors, consequently, depend on the periods of the segments,
which are easily manageable.

\subsection{Frequency-uncorrelated photon triplet}

In this subsection we shortly discuss generation of
frequency-uncorrelated triplets of photons, in analogy with the
cascaded scheme realized experimentally in \cite{Hubel}.

Thus, we assume that photon frequency-uncorrelated pairs in
three-wave cascaded process
$\omega\rightarrow\omega_{1}+\omega_{2}$,
$\omega_{2}\rightarrow\omega_{1}+\omega_{1}$ can be generated.
Note, that in typical SPDC experiments a monochromatic pump beam
is used. In this situation, the sum of the frequencies of the
signal and idler photons is fixed, thus the frequencies of photons
are anticorrelated. A number of techniques have been proposed for
creating frequency-correlated SPDC under short laser pulses
\cite{Uren}. This point is discussed in details in Sec. IV for the
cascaded scheme. In the frequency-uncorrelated approach, we can
omit the frequency dependence in the phase-mismatch functions
(\ref{zt}), (\ref{qsi}) considering them on the fixed central
frequencies $\omega, \omega_{1}$ and $\omega_{2}$ with negligible
bandwidths $\Delta\omega\ll\omega$,
$\Delta\omega_{i}\ll\omega_{i}$. In this case the effective
Hamiltonian (\ref{BHam}) is described in the standard way by
discreet creation and annihilation operators: $a^{+}, b^{+}$ and
$a, b$, respectively. The commutation relations are $[a,a^{+}]=1$,
$[b,b^{+}]=1$, but without using continuous-variable operators as
in the Hamiltonian (\ref{BHam}). For the case when photons are
spatially separated the result reads as
\begin{center}
\begin{eqnarray}
H=H_{1}+H_{2},\label{HghzT}~~~~~~~~~~~~~~~~~\\
H_{1}=i\hbar(\zeta^{'}E_{0}b^{+}a^{+}_{1}-\zeta^{'*}E^{*}_{0}ba_{1}),\label{Ham-UC1}\\
H_{2}=i\hbar(\xi^{'}a^{+}_{2}a^{+}_{3}b-\xi^{'*}a_{2}a_{3}b^{+}).\label{Ham-UC2}
\end{eqnarray}
\end{center}
Here, the subscripts label the spatial mode of photons at the
frequencies $\omega_{1}$, $E_{0}$ describes the amplitude of
driving field, $\zeta^{'}$ and $\xi^{'}$ are nonlinear coupling
constants that correspondingly coincide with $\zeta$ and $\xi$ for
fixed frequencies $\omega, \omega_{1}, \omega_{2}$, accurate
within normalization multipliers. We suppose the initial state of
the cascaded system is a vacuum state
$|\psi_{0}\rangle=|0\rangle_{b}|0\rangle_{1}|0\rangle_{2}|0\rangle_{3}$.
Thus, time evolution of the system on the formula (\ref{PsD})
leads to the final state in the following form \cite{Hubel}
\begin{equation}
|\psi(t)\rangle=-\frac{t^{2}}{\hbar^{2}}H_{2}H_{1}|\psi\rangle_{in}=E_{0}\bar{\zeta}\bar{\xi}|0_{b},1_{1},1_{2}1_{3}\rangle.\label{Ps1}
\end{equation}
Here the parameters $\bar{\zeta}=t\zeta^{'}$, $\bar{\xi}=t\xi^{'}$
describing the coupling strength between the interacting fields.

\subsection{Production of GHZ polarization entangled states}

Below we apply the results obtained for preparation of
three-photon GHZ states in the cascaded scheme. It is possible for
the case when cascaded processes involve polarized photons. Thus,
we modify the above results considering three-wave interaction in
the following form
$\sum_{\alpha,\beta,\gamma}{\chi^{\left(2\right)}_{\alpha,\beta,\gamma}(z)E_{\alpha}(z,t)E_{\beta}(z,t)}E_{\gamma}(z,t)$
(see, formula (\ref{H1})), where $\alpha,\beta,\gamma$ are the
indexes of polarization states.

Including into consideration the polarization states of the
photons we assume that the type-II process
$\omega_{0}\rightarrow\omega_{1}+\omega_{2}$ create the pair of
photons with $a$ vertical $V$ and horizontal $H$ polarizations. If
the pump field is oriented at $45^{\circ}$ to the horizontal and
vertical axes two processes
$\omega_{0}\rightarrow\omega_{1}(V)+\omega_{2}(H)$ and
$\omega_{0}\rightarrow\omega_{1}(H)+\omega_{2}(V)$ take place in
the first crystal with effective coupling constant $\chi$. The
second, type-I crystal is arranged in the manner that the
following process:
$\omega_{2}(V)\rightarrow\omega_{1}(V)+\omega_{2}(V)$ and
$\omega_{2}(H)\rightarrow\omega_{1}(H)+\omega_{2}(H)$ should be
realized (coupling constant $k$). For simplicity we will restrict
our attention considering frequency-uncorrelated three-photon
states and assume that the process under photons with $(V)$ and
$(H)$ polarizations are described by the equal coupling constants.
Thus, we assume that photon pairs in three-wave processes :
$\omega_{0}\rightarrow\omega_{1}+\omega_{2}$,
$\omega_{2}\rightarrow\omega^{'}_{1}+\omega^{"}_{1}$ have
correlations on the polarization, but not on the spectral lines.
In analogy with Eqs.(\ref{HghzT})-(\ref{Ham-UC2}), we model the
sum of the corresponding parametric interactions by the following
effective Hamiltonian
\begin{eqnarray}
H=H_{1}+H_{2},\label{ankH}~~~~~~~~~~~~~~~~\\
H_{1}=i\hbar\chi
E_{0}(a^{+}_{1}b^{+}_{2}+a^{+}_{2}b^{+}_{1})+h.c.,\\
H_{2}=i\hbar
k(b_{1}(a^{+}_{1})^{2}+b_{2}(a^{+}_{2})^{2})+h.c..\label{HGHZ}
\end{eqnarray}

Here, $a_{1}$ and $b_{1}$ are the annihilation operators at
vertical polarizations, while the operators $a_{2}$ and $b_{2}$
corresponds to the horizontal polarized photons of the frequencies
$\omega_{1}$ and $\omega_{2}$, respectively.

Time-evolution of the vector state of the system is described by
the following formula (\ref{PsD}) with the Hamiltonians
(\ref{ankH})-(\ref{HGHZ}). Choosing the initial state as a vacuum
state
$|\psi_{in}\rangle=|0\rangle_{a_{1}}|0\rangle_{a_{2}}|0\rangle_{b_{1}}|0\rangle_{b_{2}}$
for all modes we derive the final state during time evaluation in
the following form
\begin{widetext}
\begin{eqnarray}
|\psi(t)\rangle=\left(-\frac{i}{\hbar}\right)^{2}t^{2}H_{2}H_{1}|\psi\rangle_{in}=\bar{\chi}\bar{k}E_{0}\Bigg[(a^{+}_{1})^{2}a^{+}_{2}+(a^{+}_{2})^{2}a^{+}_{1}\Bigg]|0\rangle_{a_{1}}|0\rangle_{a_{2}}|0\rangle_{b_{1}}|0\rangle_{b_{2}},\label{PsGHZ}
\end{eqnarray}
\end{widetext}
where $\bar{\chi}=\chi t$ and $\bar{k}=kt$ are the coupling
constants. We generalize this results by including the subscripts
labelled the spatial modes of photons at frequency $\omega_{1}$
(see, Eqs. (\ref{Ham-UC1}),(\ref{Ham-UC2}),(\ref{Ps1}). Then, the
prepared three-photon polarization-entangled states can be written
as
\begin{equation}
|\psi(t)\rangle=\bar{\chi}\bar{k}E_{0}\left(|V\rangle|H\rangle|H\rangle+|H\rangle|V\rangle|V\rangle\right)|0\rangle_{b_{1}}|0\rangle_{b_{2}},
\end{equation}
where the states $|V\rangle=a^{+}_{1}|0\rangle_{a_{1}}$,
$|H\rangle=a^{+}_{2}|0\rangle_{a_{2}}$ present the vertical and
horizontal polarization states of photons at the frequency
$\omega_{1}=\frac{\omega_{0}}{3}$. Thus, we demonstrate that in
this cascaded scheme triple photons can constitute the
Greenberger-Horne-Zeilinger states of light.

\section{QPM in dual section nonlinear crystal with linear elements}

\begin{figure}
\includegraphics[width=8cm]{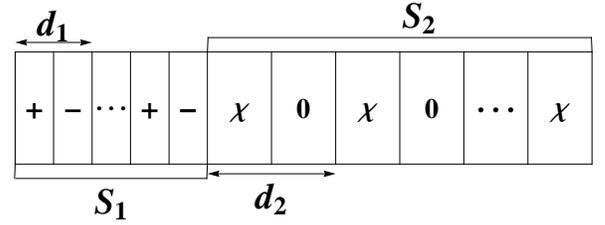}
\caption{Scheme of dual section layered structure, $S_{1}$ is a
first section with the length $L_{1}$, which involves  nonlinear
domains of length $l_{1}$ with $\chi_{+}>0$ and $\chi_{-}<0$ (with
period $d_{1}$) and $S_{2}$ is the second section with the length
$L_{2}$, that involves both nonlinear layers of length $l_{2}$
with $\chi$ susceptibilities of the same sign and linear
dispersive optical spacers of length $l_{3}$ (period $d_{2}$).}
\label{L2sec}
\end{figure}

In this section we investigate an improvement of above approach
based on dual-grid nonlinear crystal for engineering controllable,
factorable photon states. We consider multilayered structure of
second-order nonlinear and linear materials. The key of this
investigation is the idea of manipulating overall group velocity
delay mismatches between the various fields in structured media
for engineering twin-photon states with special temporal and
spectral characteristics from. This approach is expanded here for
the case of multistep parametric processes SPDC \cite{WSAK}. We
analyze the dual-grid structures to mismatch two processes. This
method has been recently used in arranging of cascaded OPO
\cite{Kolk}. The modified crystal consists of two family of
segments: $M$ segments of the length $l_{1}$, with positive $\chi$
and negative $-\chi$ susceptibilities (corresponding
phase-matching function is $\Delta k_{1}$), and the second is the
structure, which consists of $\frac{N}{2}$ segments of the length
$l_{2}$ with  $\chi$ susceptibilities and $\frac{N}{2}$ linear
optical spacers of the length $l_{3}$ (corresponding two
phase-matching functions are $\Delta k_{2}$ and $\Delta \kappa$
for the nonlinear and the linear segments). The $\omega_{0}
\rightarrow {\omega_{1}+\omega_{2}}$ and $\omega_{2} \rightarrow
{\omega_{1}+\omega_{1}}$ processes have been implemented in the
first and the second sections of the crystal, correspondingly
(see, Fig.~\ref{L2sec}).

We calculate the coupling constants of parametric processes:
$\zeta$ and $\xi$ as a function of the finite number of nonlinear
and linear segments. We again assume that effective $\omega_{0}
\rightarrow {\omega_{1}+\omega_{2}}$ process goes in the first
section, so the conditions of the perfect QPM is $\Delta k_{1}=q$
($q=\frac{2\pi}{d_{1}}$). Thus, the result for $\zeta$ is exactly
the same as the first term in (\ref{SumZt-Hum}).

The Fourier function (\ref{qsi}) is calculated in the following
form
\begin{eqnarray}
\xi= l_{2}\chi
e^{-i\phi_{2}}sinc\left(\frac{l_{2}}{2}\Delta{k_{2}}\right)
\frac{\sin{\left(\frac{N\left(l_{2}+l_{3}\right)\Delta{K}}{4}\right)}}{\sin{\left(\frac{\left(l_{2}+l_{3}\right)\Delta{K}}{4}
\right)}},  \label{SumXi}
\end{eqnarray}
where $d_{2}=\left(l_{2}+l_{3}\right)$ is a period of the
 second section, $\phi_{2}=\frac{1}{2}l_{2}\Delta{k_{2}}+\Delta
 K(l_{2}+l_{3})(\frac{N-1}{2})$ and
$L_{2}=\frac{N}{2}\left(l_{2}+l_{3}\right)$ is the total length of
the second section, where goes $\omega_{2} \rightarrow
{\omega_{1}+\omega_{1}}$ process with the following phase-matching
functions (\ref{k1})
\begin{eqnarray}
\Delta k_{2}= k_{2}(\omega_{2})-k_{1}(\omega^{'}_{1})-k_{1}(\omega^{"}_{1}),\nonumber\\
\Delta \kappa_{2}= \kappa_{2}(\omega_{2})-\kappa_{1}(\omega^{'}_{1})-\kappa_{1}(\omega^{"}_{1}),\nonumber\\
\Delta{K}=\bar{l_{2}}\Delta{k_{2}}+\bar{l_{3}}\Delta{\kappa_{2}},
\end{eqnarray}
 where $\bar{l_{i}}=\frac{l_{i}}{l_{2}+l_{3}}$,
 $\left(i=2,3\right)$,  $\bar{l_{2}}+\bar{l_{3}}=1$.
The condition of perfect QPM in this configuration is $\Delta
K=0$. We represent the phase-matching functions of nonlinear:
$\Delta k_{2}$ and linear: $\Delta \kappa$ segments as following
$\Delta k_{2}=\Delta k^{\left(0\right)}_{2}+\delta k_{2}$ and
$\Delta \kappa_{2}=\Delta \kappa^{\left(0\right)}_{2}+\delta
\kappa_{2}$. We assume that each domain is aligned such that
zero-order  constant of nonlinear and linear terms compensate each
other $l_{2}\Delta k^{\left(0\right)}_{2}+l_{3}\Delta
\kappa^{\left(0\right)}_{2}=0$. The condition of perfect QPM
requires the quantities $\delta k_{2}$ and $\delta \kappa$ exhibit
opposite signs: $\bar{l}_{2}\delta k_{2}=-\bar{l}_{3}\delta
\kappa$. We additionally require the following condition $\Delta
\kappa^{\left(0\right)}_{2}=\frac{\pi}{l_{3}}$ for zero-order
therm of phase-matching function of linear segments. These
conditions are realistic and easily satisfying by managing the
lengths of the nonlinear and linear segments, and managing the
dispersive parameters of linear segments during crystal growth.
The perfect QPM of both processes brings to maxima of coupling
constants of the corresponding processes which leads to excellent
and effective three-photon down-conversion.

\subsection{Factorable three-photon states. Heralded two-photon polarization states}

Pure single photon states are probably the most fundamental
entities in quantum optics, and constitute the   starting point
for many optically-based quantum enhanced technologies. A basic
requirement for many key applications is the ability to generate
reliably pure single photon wavepackets capable of high-visibility
interference. As we noted, single photon wavepackets may be
generated from number-correlated pairs by conditional state
preparation. For example, photon pairs produced by the process of
spontaneous parametric down-conversion (PDC) in a
$\chi^{\left(2\right)}$ nonlinear medium allow conditional
preparation of spectrally factorized photon pairs. The generated
pure photons remain polarization entangled, thus, the detection of
one photon in the pair heralds the presence of the conjugate
photon.

Below we are going to analyze deduced results for the three-photon
amplitude of down-converted light for the cascaded processes
$\omega_{0}\rightarrow\omega_{2}+\omega_{1}$ and
$\omega_{2}\rightarrow\omega^{'}_{1}+\omega^{"}_{1}$ in the
dual-section crystal (see, Fig.~\ref{L2sec}) and find conditions
for generation spectrally factorized photon states. Each of the
generated modes has own polarization and we take into account
their polarizations in the coupling constants $\xi$ and
$\varsigma$ and hence in the amplitude
$\Phi(\omega_{1},\omega^{'}_{1},\omega^{"}_{1})$ according to the
remark presented after formula (\ref{klishkoEq}). Thus, we take
into account the polarization indexes $\alpha,\beta,\gamma=o,e$ in
the vector waves of the modes. To find the conditions of
factorization we need to write the three-photon amplitude from
Eq.(\ref{FipD}) in the Gaussian form. For this reason we expand
all phase mismatching vectors in Taylor series and take only the
zero- and first-order terms.
\begin{eqnarray}
l_{1}\Delta k_{1}=l_{1}\Delta
k^{\left(0\right)}_{1}+T_{1,\alpha}\nu_{1}+T_{2,\beta}\nu_{2}+T_{3,\gamma}\nu_{3},\\
l_{2}\Delta k_{2}=l_{2}\Delta
k^{\left(0\right)}_{2}+t_{2,\beta}\nu_{2}+t_{3,\gamma}\nu_{3},\\
l_{3}\Delta \kappa=l_{3}\Delta
\kappa^{\left(0\right)}_{2}+\varrho_{2,\beta}\nu_{2}+\varrho_{3,\gamma}\nu_{3}.
\end{eqnarray}
Here: \(\nu_{0}=\omega_{0},
\nu_{1}=\omega_{1}-\frac{\omega_{0}}{3}\),
\(\nu_{2}=\omega^{'}_{1}-\frac{\omega_{0}}{3}\),
\(\nu_{3}=\omega^{"}_{1}-\frac{\omega_{0}}{3}\),
\(\nu=\omega_{2}-\frac{\omega_{0}}{3}\). The coefficients
presented above have the following definitions:

\begin{eqnarray}
T_{i,\alpha}=l_{1}\left(\frac{d
k_{L}\left(\nu_{0},\beta\right)}{d\nu_{0}}|_{\nu_{0}=\omega_{0}}-\frac{d
k_{1}\left(\nu_{i},\alpha\right)}{d\nu_{i}}|_{\nu_{i}=0}\right)=\nonumber\\
l_{1}\left(u^{-1}_{L,\beta}-u^{-1}_{i,\alpha}\right), \label{T}
\end{eqnarray}
where $i=1,2,3$, $\alpha,\beta=o,e$, for the arbitrary
polarization of the laser field;
\begin{eqnarray}
t_{\mu,\gamma}=l_{2}\left(\frac{d
k_{2}\left(\nu,\beta\right)}{d\nu}|_{\nu=\frac{\omega_{0}}{3}}-\frac{d
k_{1}\left(\nu_{\mu},\gamma\right)}{d\nu_{\mu}}|_{\nu_{\mu}=0}\right)=\nonumber\\
l_{2}\left(u^{-1}_{\beta}-u^{-1}_{\mu,\gamma}\right), \label{t}
\end{eqnarray}
for arbitrary $\beta=o,e$ polarization of  the mode with the
central frequency $\frac{2\omega_{0}}{3}$ in the nonlinear
domains;
\begin{eqnarray}
\varrho_{\mu,\gamma}=l_{3}\left(\frac{d
\kappa_{2}\left(\nu,\beta\right)}{d\nu}|_{\nu=\frac{\omega_{0}}{3}}-\frac{d
\kappa_{1}\left(\nu_{\mu},\gamma\right)}{d\nu_{\mu}}|_{\nu_{\mu}=0}\right)=\nonumber\\
l_{3}\left(\upsilon^{-1}_{\beta}-\upsilon^{-1}_{\mu,\gamma}\right),
\label{Ro}
\end{eqnarray}
where $\mu=2,3$, $\beta,\gamma=o,e$, the $\beta$ polarization of
the mode with the central frequency $\frac{2\omega_{0}}{3}$ in the
linear domains is arbitrary.

Here, $T_{i,\alpha}$ are the temporal walkoffs between pumped and
down-converted pulses (with the corresponding group velocities
$u^{-1}_{L,\beta}$ and $u^{-1}_{\mu,\alpha}$); $t_{\mu,\gamma}$
and $\varrho_{\mu,\gamma}$ are the temporal walkoffs between modes
with the central frequency $\omega_{2}$ and down-converted pulses
with the cental frequency $\omega_{1}$ for nonlinear and linear
domains, correspondingly (with the group velocities in nonlinear
$u^{-1}_{\beta}, u^{-1}_{\mu,\gamma}$ and linear
$\upsilon^{-1}_{\beta}, \upsilon^{-1}_{\mu,\gamma}$ domains. Here
$u^{-1}_{\beta}, \upsilon^{-1}_{\beta}$ and $u^{-1}_{\mu,\gamma},
\upsilon^{-1}_{\mu,\gamma}$ are the group velocities of modes at
frequencies $\omega_{2}$ and $\omega_{1}$, respectively).

Taking into account that the pump pulse has Gaussian form
(\ref{EG}) and using the formulas (\ref{SumZt-Hum}), (\ref{SumXi})
and (\ref{T})-(\ref{Ro}) we find the following form for the
three-photon amplitude

\begin{widetext}
\begin{eqnarray}
\Phi_{\alpha,\beta,\gamma}(\nu_{1},\nu_{2},\nu_{3})=F_{\alpha,\beta,\gamma}(\nu_{1},\nu_{2},\nu_{3})f_{\alpha}(\nu_{1})f_{\beta}(\nu_{2})f_{\gamma}(\nu_{3})\Phi_{\alpha,\beta}(\nu_{1},\nu_{2})\Phi_{\alpha,\gamma}(\nu_{1},\nu_{3})\Phi_{\beta,\gamma}(\nu_{2},\nu_{3}),\label{FiGaus}
\label{Fi3}
\end{eqnarray}
\end{widetext}

\begin{widetext}
\begin{eqnarray}
F_{\alpha,\beta,\gamma}(\nu_{1},\nu_{2},\nu_{3})=\exp\Bigg(-\frac{iM}{2}(T_{1,\alpha}\nu_{1}+T_{2,\beta}\nu_{2}+T_{3,\gamma}\nu_{3})\Bigg)\exp\Bigg(-\frac{iN}{4}(t_{2,\beta}\nu_{2}+t_{3,\gamma}\nu_{3}+\varrho_{2,\beta}\nu_{2}+\varrho_{3,\gamma}\nu_{3})\Bigg),
\end{eqnarray}
\end{widetext}

\begin{widetext}
\begin{eqnarray}
f_{\alpha}(\nu_{1})=\exp\Bigg(-\left(\frac{\tau^{2}_{p}}{2}+\frac{M^{2}}{20}T^{2}_{1,\alpha}\right)\nu^{2}_{1}\Bigg),\\
f_{\beta}(\nu_{2})=\exp\Bigg(-\left(\frac{\tau^{2}_{p}}{2}+\frac{M^{2}}{20}T^{2}_{2,\beta}+\frac{N^{2}}{80}(t_{2,\beta}+\varrho_{2,\beta})^{2}\right)\nu^{2}_{2}\Bigg),\\
f_{\gamma}(\nu_{3})=\exp\Bigg(-\left(\frac{\tau^{2}_{p}}{2}+\frac{M^{2}}{20}T^{2}_{3,\gamma}+\frac{N^{2}}{80}(t_{3,\gamma}+\varrho_{3,\gamma})^{2}\right)\nu^{2}_{3}\Bigg),
\end{eqnarray}
\begin{eqnarray}
\Phi_{\alpha,\beta}(\nu_{1},\nu_{2})=\exp\Bigg(-\left[\tau^{2}_{p}+\frac{M^{2}}{10}T_{1,\alpha}T_{2,\beta}\right]\nu_{1}\nu_{2}\Bigg),\\
\Phi_{\alpha,\gamma}(\nu_{1},\nu_{3})=\exp\Bigg(-\left[\tau^{2}_{p}+\frac{M^{2}}{10}T_{1,\alpha}T_{3,\gamma}\right]\nu_{1}\nu_{3}\Bigg),\\
\Phi_{\beta,\gamma}(\nu_{2},\nu_{3})=\exp\Bigg(-\left[\tau^{2}_{p}+\frac{M^{2}+1}{10}T_{2,\alpha}T_{3,\beta}+\frac{N^{2}}{40}(t_{2,\beta}+\varrho
_{2,\beta})(t_{3,\gamma}+\varrho
_{3,\gamma})\right]\nu_{2}\nu_{3}\Bigg).
\end{eqnarray}
\end{widetext}

Let us specify the three-photon process of interest and write
conditions for spectral factorization. We concentrate on the
discussion of $eoo$ state. It means that in the first cascaded
process: $\omega_{0}\rightarrow\omega_{2}+\omega_{1}$ photon with
central frequency $\frac{\omega_{0}}{3}$ and polarization $e$ is
generated, while in the second process:
$\omega_{2}\rightarrow\omega^{'}_{1}+\omega^{"}_{1}$ two photons
with central frequencies $\frac{\omega_{0}}{3}$ and polarizations
$o$ are generated. All three photons are strongly correlated, as
they are generated from one input photon. Thus, the photon with
polarization $e$ can be used as a heralded photon. The condition
for generation of correlated photon pairs and the heralded photon
is
\begin{equation}
|\Phi_{e,o,o}(\nu_{1},\nu_{2},\nu_{3})|^{2}=|\Phi^{\left(1\right)}_{e}(\nu_{1})|^{2}|\Phi^{\left(2\right)}_{o,o}(\nu_{2},\nu_{3})|^{2},\label{Fc2}
\end{equation}
where the separate factor
$|\Phi^{\left(1\right)}_{e}(\nu_{1})|^{2}=|f_{e}(\nu_{1})|^{2}$
describes one-photon probability and
$|\Phi^{\left(2\right)}_{o,,o}(\nu_{2},\nu_{3})|^{2}=|f_{o}(\nu_{2})\Phi_{o,o}(\nu_{2},\nu_{3})f_{o}(\nu_{3})|^{2}$
describes two-photon probability.

Here are represented the conditions which have to be satisfied for
generation of heralded photon
\begin{eqnarray}
\tau^{2}_{p}+\frac{M^{2}}{10}T_{1,e}T_{2,o}=0,\nonumber\\
\tau^{2}_{p}+\frac{M^{2}}{10}T_{1,e}T_{3,o}=0.\label{F2G}
\end{eqnarray}
These conditions are reduced to one because as it is seen from
Eq.(\ref{T}) $T_{2,o}=T_{3,o}$. We found these conditions from the
idea that the joint three-photon state in the Gaussian form
(\ref{FiGaus}) must be free of correlations between
$\nu_{1},\nu_{2}$ and $\nu_{1},\nu_{3}$ spectral components.

Note, that each of conditions (\ref{F2G}) in form coincides with
the condition derived for the spectral factorization of two-photon
amplitude in down-conversion  \cite{Uren,KZ95}. It is seen from
Eq.(\ref{T}), that the explanation of this results reveals, that
this condition may be satisfied either when
$u^{-1}_{2,o}<u^{-1}_{L\beta}<u^{-1}_{1,e}$ or when
$u^{-1}_{1,e}<u^{-1}_{L,\beta}<u^{-1}_{2,o}$, (for arbitrary
polarization of the laser field $\beta$). It means that the group
velocity of pump pulse ought to lie between the group velocities
of the generated pulses with different polarizations. This method
provides the optimal compensation for the walkoff effect
\cite{KZ95,KelRub}. It should be point out that it is impossible
for type-I down conversion, since in this process photons with
similar polarization are produced, consequently, with the same
group velocities. We could manage the duration of the pump field
$\tau_{p}$ and the length of the domain $l_{1}$ for perfect
compensation of the terms in the conditions.

In the result of cascaded three-photon down conversion we could
generate completely spectrally factorized pure tree photon states:
$|\Phi_{e,o,o}(\nu_{1},\nu_{2},\nu_{3})|^{2}=|\Phi^{\left(1\right)}_{e}(\nu_{1})|^{2}|\Phi^{\left(1\right)}_{o}(\nu_{2})|^{2}|\Phi^{\left(1\right)}_{o}(\nu_{3})|^{2}$.
The conditions we have to satisfy are Eqs.(\ref{F2G}) complemented
by the following condition

\begin{equation}
\tau^{2}_{p}+\frac{M^{2}}{10}T_{2,e}T_{3,o}+\frac{N^{2}}{40}(t_{2,o}+\varrho
_{2,o})(t_{3,o}+\varrho _{3,o})=0.\label{F3C}
\end{equation}
The last is found from the requirement that the joint three-photon
state additionally must be free of the correlations between
$\nu_{2},\nu_{3}$ spectral components. We have observed the
dual-section composite crystal with linear elements, which helped
us to achieve perfect quasi-phase-matching for two simultaneous
processes. The linear elements play additional positive role in
the compensation of the terms in Eq.(\ref{F3C}), thus, in the
factorization of the states. In such configurations, there are
several parameters which can be managed easily: lengths of the
nonlinear $l_{1}, l_{2}$ and linear $l_{3}$ domains, number of
linear and nonlinear domains with different lengths, the bandwidth
of the pump pulse $\tau_{p}$, as well as the dispersion
coefficients of linear domains $\varrho_{\mu}$. The condition
(\ref{F3C}) is easy achievable, if the dispersion coefficients of
the fields in the linear spacers $\varrho_{\mu}$ have the opposite
sign to those of in the nonlinear domains $t_{\mu}$. Thus, we
could arrange the lengths of nonlinear and linear domains in the
second segment, as well as the numbers of domains such that the
above condition (\ref{F3C}) is fulfilled.

Thus, we have demonstrated the modification of QPM with linear
compensator \cite{Uren,Kly} for the cascaded processes. It is
shown that this scheme enables the engineering triplet of photons
with desirable spectral properties. Using the feature of
polarization entanglement, it is possible to consider one of the
photons as heralded photon and operate with two renaming states
definitely knowing their polarizations.

\section{Three-photon correlation and Wigner functions in Cascaded OPO}

In the presence of the optical cavity the cascaded three-photon
splitting displays new properties that are the subject of this
section. We assume that the cavity supports three modes at
frequencies $\omega_{0}$, $\omega_{1}=\frac{\omega_{0}}{3}$ and
$\omega_{2}=\frac{2\omega_{0}}{3}$. The mode $\omega_{0}$ is the
pump mode, steerable by an external coherent driving field at the
same frequency $\omega_{0}$, while the modes $\omega_{1}$ and
$\omega_{2}$ are the modes of the generating subharmonics. In this
case, the cavity modes are described by discreet creation and
annihilation operators $a^{+}, b^{+}$ and $a, b$ as in the Sec.
II, III.

\begin{figure*}
\begin{center}
\includegraphics[width=14cm]{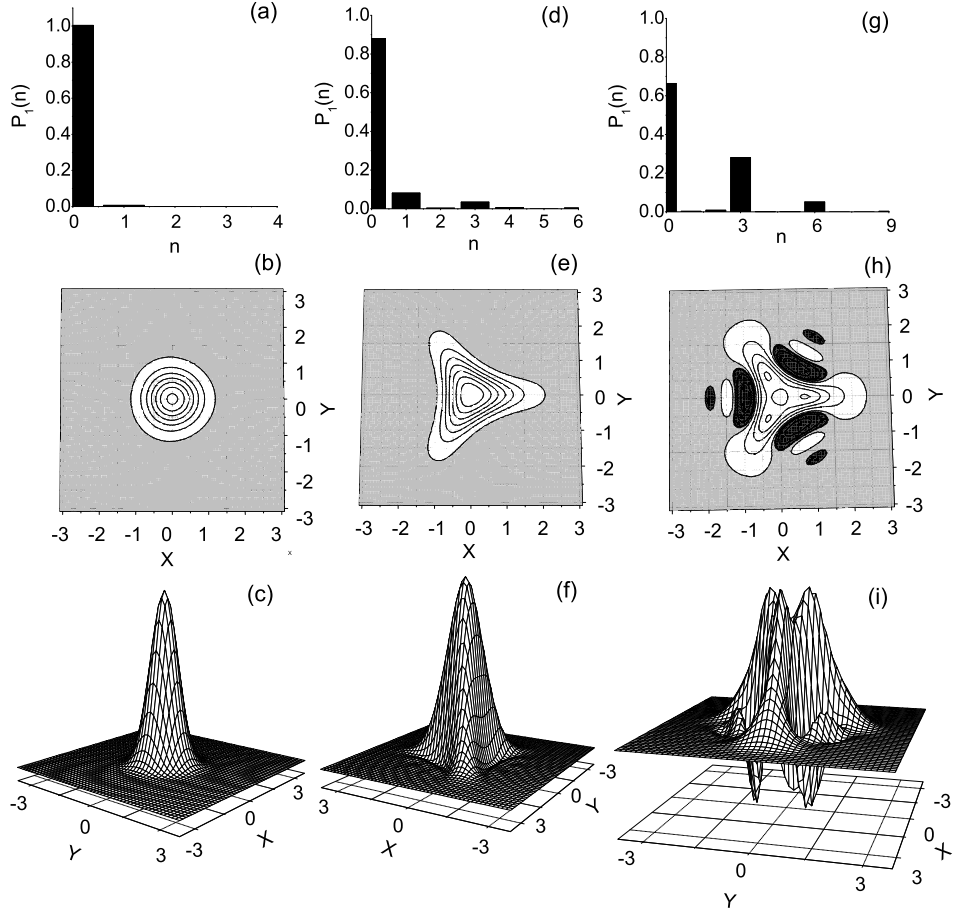}
\end{center}
\caption{Photon number distributions (a, d, g), the Wigner
functions (c, f, i) and its contour plots (b, e, h) for
$\omega_{1}$-mode and for different time intervals, $t= 5\times
10^{-4}\gamma^{-1}$ (a, b, c); $t= 5\times 10^{-3}\gamma^{-1}$ (d,
e, f); $t= 2\times 10^{-2}\gamma^{-1}$ (g, h, i). The other
parameters are: $\zeta^{'}/\gamma=200$, $\xi^{'}/\gamma=100$,
$\gamma_{2} = \gamma_{1}=\gamma$.}\label{9W}
\end{figure*}

The model Hamiltonian for the system, in the rotating-wave
approximation, is given by
\begin{eqnarray}
H_{int}=i\hbar\left(Ee^{-i\omega_{0}t}a^{+}_{0}-E^{*}e^{i\omega_{0}t}a_{0}\right)+\nonumber\\
i\hbar\zeta^{'}\left(a_{0}a^{+} b^{+}-a^{+}_{0}a b
\right)+i\hbar\xi^{'}\left(a^{+2}
b-a^{2}b^{+}\right).\label{Hamint}
\end{eqnarray}
Here, $E$ describes the amplitude of the driving fields,
$\zeta^{'}$ and $\xi^{'}$ are the nonlinear coupling constants
that are proportional to the functions $\zeta$ and $\xi$
(\ref{zt}), (\ref{qsi}), correspondingly. In difference from the
Hamiltonian (\ref{HghzT}) this one includes the first term that describes
excitation of the pump mode $\omega_{0}$, by a classical pump
field at the same frequency.

The cascaded OPO is dissipative, because the modes suffer from
losses due to partially transmission of light through the mirrors
of the cavity. The master equation for the density operator $\rho$
of the cavity modes in the Limbland form reads as
\begin{equation}
\frac{\partial\rho}{\partial
t}=\frac{1}{i\hbar}[H_{int}\rho]+\sum_{i=0,1,2}{\gamma_{i}\left(2a_{i}\rho
a^{+}_{i}-a^{+}_{i}a_{i}-\rho a^{+}_{i}a_{i}\right)},\label{dRo}
\end{equation}
where $\gamma_{i}$ are the cavity damping rates for the modes
$\omega_{i}$, ($i=0,1,2$).

Below, we briefly review the semiclassical results on the model
following the papers \cite{y, Kryuchkyan2}. For the sake of
simplicity let us first of all consider the case of high cavity
losses for the fundamental mode
$\gamma_{0}\gg\gamma_{1},\gamma_{2}$, assuming the pump depletion
and, hence, its back influence on the modes (1) and (2) are
completely neglected. Then the steady state solutions, for
$t\gg\gamma^{-1}_{1},\gamma^{-1}_{2}$, is founded in above
threshold operational regime $\varepsilon=\frac{E}{E_{th}}>1$,
where
$E_{th}=\frac{2\gamma_{0}}{3\zeta^{'}}\sqrt{2\gamma_{1}\gamma_{2}}$.
The mean photon numbers of the modes are
\begin{eqnarray}
n_{1}=\frac{\gamma_{1}\gamma_{2}}{18\xi^{'2}}\left(\varepsilon+3\sqrt{\varepsilon^{2}-1}\right)^{2}, \label{n1}\\
n_{2}=\frac{\gamma^{2}_{1}}{36\xi^{'2}}\left(\frac{\varepsilon+3\sqrt{\varepsilon^{2}-1}}{\varepsilon+\sqrt{\varepsilon^{2}-1}}\right)^{2},
\label{n2}
\end{eqnarray}
and the phases are equal to
\begin{equation}
\phi_{1}=\frac{\Phi}{3}+\frac{2\pi}{3}n,~~\phi_{2}=\frac{\Phi}{3}-\frac{2\pi}{3}n,~~
n=0,1,2,\label{FiS}
\end{equation}
where $\Phi$ is the phase of the driving field $E=|E|e^{i\Phi}$.

The stability analysis shows, that solutions (\ref{n1}),(\ref{n2})
for both subharmonic modes are stable in the range $E>E_{th}$.
However, a bistable hysteresis-cycle behavior of the $n_{1}$ and
$n_{2}$, in dependence on the amplitude $E$, is realized in a
small interval $1<\varepsilon<\frac{3}{2\sqrt{2}}$. Note, that
analysis of semiclassical solutions involving the depletion
effects has been done in \cite{Kryuchkyan2}.

The second part of this section is devoted to the study of quantum
statistical properties of high-intensity subharmonic mode
$\omega_{1}$ including calculations of the corresponding Wigner
functions and correlation function. Calculations are performed on
the framework of the quantum trajectory simulation method
\cite{Gisin}. Some details of quantum states diffusion method
(QSD), in application to study nonlinear processes in a cavity,
can be found in the papers \cite{AMK1, AMK2}.

\subsection{Photon-number distributions and Wigner functions for photon triplets.}

It seems that the correlation between photons in the triplet, as
well as quantum interference effects can be evidently displayed
for short interaction time intervals much shorter than the
characteristics relaxation time, $t \ll \gamma^{-1}$. We
illustrate these effects numerically on the base of the master
equation, however, in the regimes when the dissipation in the
cavity is unessential and the dynamic of modes is almost unitary.
For the cavity configuration presented the validity of such
approximation is guaranteed by consideration of short interaction
time $1/ \zeta^{'}, 1/ \xi^{'} \ll t \ll 1/ \gamma_{1, 2}$
provided that the coupling constants $ \zeta^{'} $ and  $ \xi^{'}
$ exceed the dumping rates for the modes.

The reduced density operators for each of the modes are
constructed from the density operator of the both modes $\rho$ by
tracing over the other mode: $\rho_{1(2)}=Tr_{2(1)}(\rho)$. The
photon number distribution for $\omega_{1}$ mode is calculated as
the diagonal elements $P_{1}(n)= \langle n| \rho_{1} | n \rangle $
on the photon-number states $| n \rangle $, while calculations of
the Wigner function for $\omega_{1}$ mode are performed by using
its standard form in a Fock space

\begin{equation}
W_{1}(\rho, \theta) = \sum_{mn}\rho_{1, mn}W_{mn}(\rho, \theta).
\end{equation}

Here: $\rho, \theta$ are the polar coordinates in the complex
phase-space, which is determined by position and momentum of the
quadratures $x=(a+a^{+})/ \sqrt{2}, y=(a-a^{+})/ \sqrt{2}i  $,
respectively, while the coefficients $W_{mn}(\rho, \theta)$ are
the Fourier transforms of matrix elements of the Wigner
characteristic function.

Examples of both photon-number distribution functions and the
Wigner functions for $\omega_{1}$-mode are plotted in Fig.
\ref{9W} for the different short time intervals. In Figs. \ref{9W}
(a),(b),(c) we depict the results for $t= 5\times
10^{-4}\gamma^{-1}$. Here, the $\omega_{1}$-mode is in the vacuum
state, therefore the corresponding Wigner function is Gaussian.
After evolution of the system, for the  time $t= 5\times
10^{-3}\gamma^{-1}$ one-photon state at the frequency $\omega_{1}$
is appeared due to the process $\omega_{0} \rightarrow \omega_{1}
+ \omega_{2} $, then photon triplet is generated due to the
cascading $\omega_{0} \rightarrow \omega_{1} + \omega_{2}
\rightarrow 3\omega_{1}$ (see, photon distribution on the Fig.
\ref{9W} (d)). The three developing arms on the Wigner function
follow the semiclassical directions (see, formula (\ref{FiS})) of
the phase space (see, Fig. \ref{9W}(e), (f)). With increasing time
intervals $t= 2\times 10^{-2}\gamma^{-1}$ pronounced three-photon
structure of the $\omega_{1}$ mode is displayed on Fig.
\ref{9W}(g), (h), (i). As we see, in this case, the most probable
values of photon numbers are separated by three photons. The
Wigner function shows three phase-components with an interference
pattern in the regions between them. We show the regions of
quantum interference in the contour plot (see, Fig. \ref{9W}(h))
depicting negative regions of the interference terms in the black.
Note, that three-fold symmetry of the Wigner function and
interference pattern have been demonstrated for the direct
three-photon down-conversion in $\chi^{\left(3\right)}$ media
\cite{Banaszek}. However, we note that the results presented here
for the cascaded configuration are different on the details from
the analogous calculation on the Wigner function in
\cite{Banaszek}.

Below, we consider photon-number distributions, as well as the
Wigner functions of the both modes at the frequencies
$\omega_{1}=\frac{\omega_{0}}{3}$ and
$\omega_{2}=\frac{2\omega_{0}}{3}$ in over-transient regime of
OPO, when dissipative effect should be included.

\subsection{Photon-number correlations.}

In the paper \cite{Hubel} the experimental verification of
time-correlation between photons in triplet has been demonstrated.
In this section, we study normalized third-order correlation
function $g^{\left(3\right)}$ of the photon-numbers of mode
$\omega_{1}$. This investigation is complemented by the
consideration of photon number distributions of subharmonic modes.
Thus, the correlation function for zero-delay time intervals is
expressed as
\begin{figure}
\includegraphics[width=7.6cm]{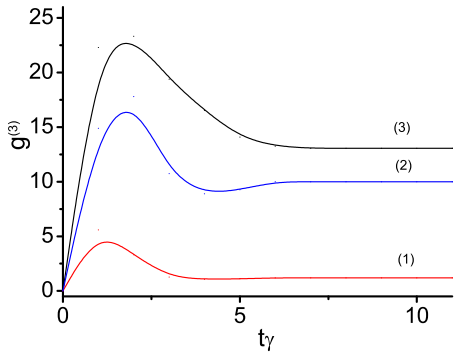}
\caption{Third-order correlation function versus $t\gamma$  on the
threshold $E/E_{th}=1$, curve (1); below the threshold
$E/E_{th}=0.7$, curve (2) and $E/E_{th}=0.4$, curve (3). The
parameters are: $\zeta^{'}/\gamma=0.2$, $\xi^{'}/\gamma=0.1$,
$\gamma_{2} = \gamma_{1}=\gamma$.} \label{G3}
\end{figure}

\begin{equation}
g^{\left(3\right)}(t)=\frac{Tr[a^{+3}a^{3}\rho_{1}(t)]}{n^{3}_{1}(t)},
\end{equation}
where $n_{1}(t)=Tr(a^{+}a\rho_{1}(t))$ is the mean photon number.
We calculate this quantity in transition near to the generation
threshold of OPO in the over transient regime, $t\gg
\gamma^{-1}_{1},\gamma^{-1}_{2}$. The results of numerical
calculations, on the base of the master equation (\ref{dRo}) with
the Hamiltonian (\ref{Hamint}), are shown in Fig.\ref{G3} for
three operation regimes of OPO. As we see, the curves (2) and (3)
describe the strong correlation between three photons registered
at the same time-intervals, that lead to three-photon
superbunching $ g^{(3)}\gg 1$. Such superbunching effect usually
takes place for the direct three-photon down-conversion. We
demonstrate analogous correlation for photon triplet \cite{Hubel}
generated in the cascaded scheme. This effect decreases if the
system moves to the range of generation threshold. At the
threshold we have $g^{\left(3\right)}=1.2$, for $t\gamma\gg 1$
(see, curve (1), Fig. 3).

\begin{figure}
\includegraphics[width=7cm]{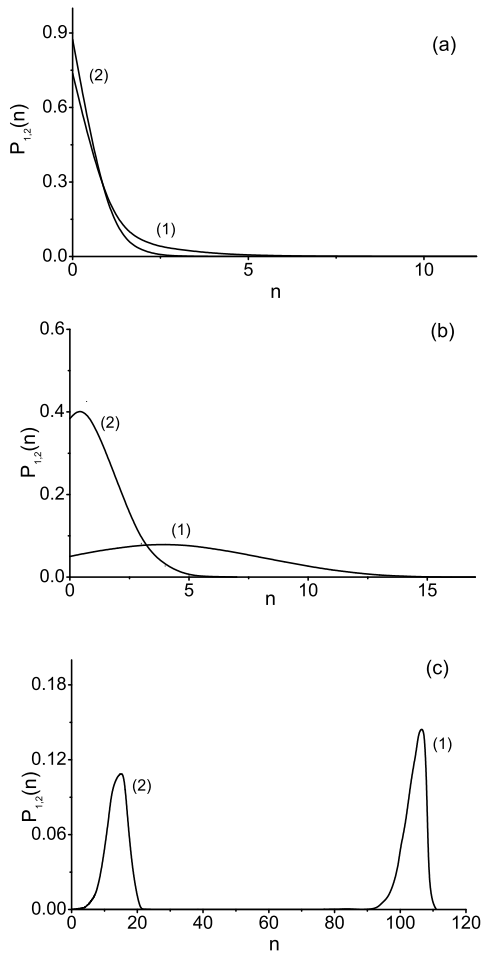}
\caption{Photon number distributions in transition through the
generation threshold $E/E_{th}=0.7$ for the mode (1) at the
frequency $\omega_{1}=\omega_{0} / 3$ (curves (1)) and mode (2) at
the frequency $\omega_{2}= 2\omega_{0} /{3}$ (curves (2)). The
regimes are: (a); $E/E_{th}=1$ (b); $E/E_{th}=1.4$ (c) for the
parameters: $\zeta^{'}/\gamma=0.2$, $\xi^{'}/\gamma=0.1$,
$\gamma_{2}=\gamma_{1}=\gamma$. } \label{Pns}
\end{figure}

\begin{figure}
\includegraphics[width=8.2cm]{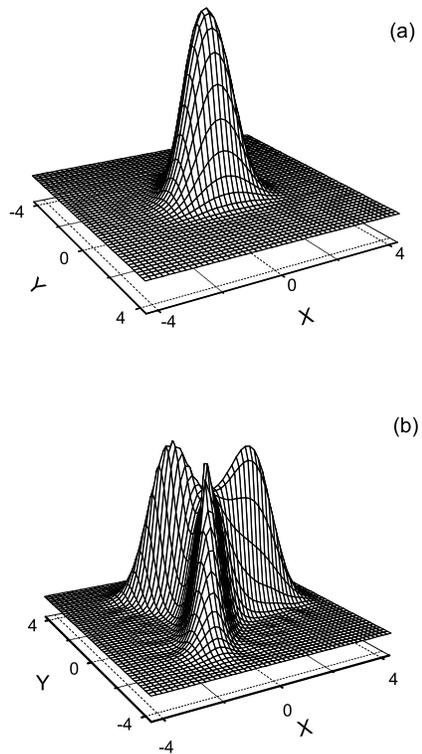}
\caption{Wigner functions for the modes $\omega_{2}= 2\omega_{0}
/{3}$ (a) and $\omega_{1}= \omega_{0} /{3}$ (b) at the generation
threshold for the parameters: $\zeta^{'}/\gamma=0.2$,
$\xi^{'}/\gamma=0.1$, $\gamma_{2}=\gamma_{1}=\gamma$ and
$E/E_{th}=1$.}\label{Wig07}
\end{figure}

\begin{figure}
\includegraphics[width=7.2cm]{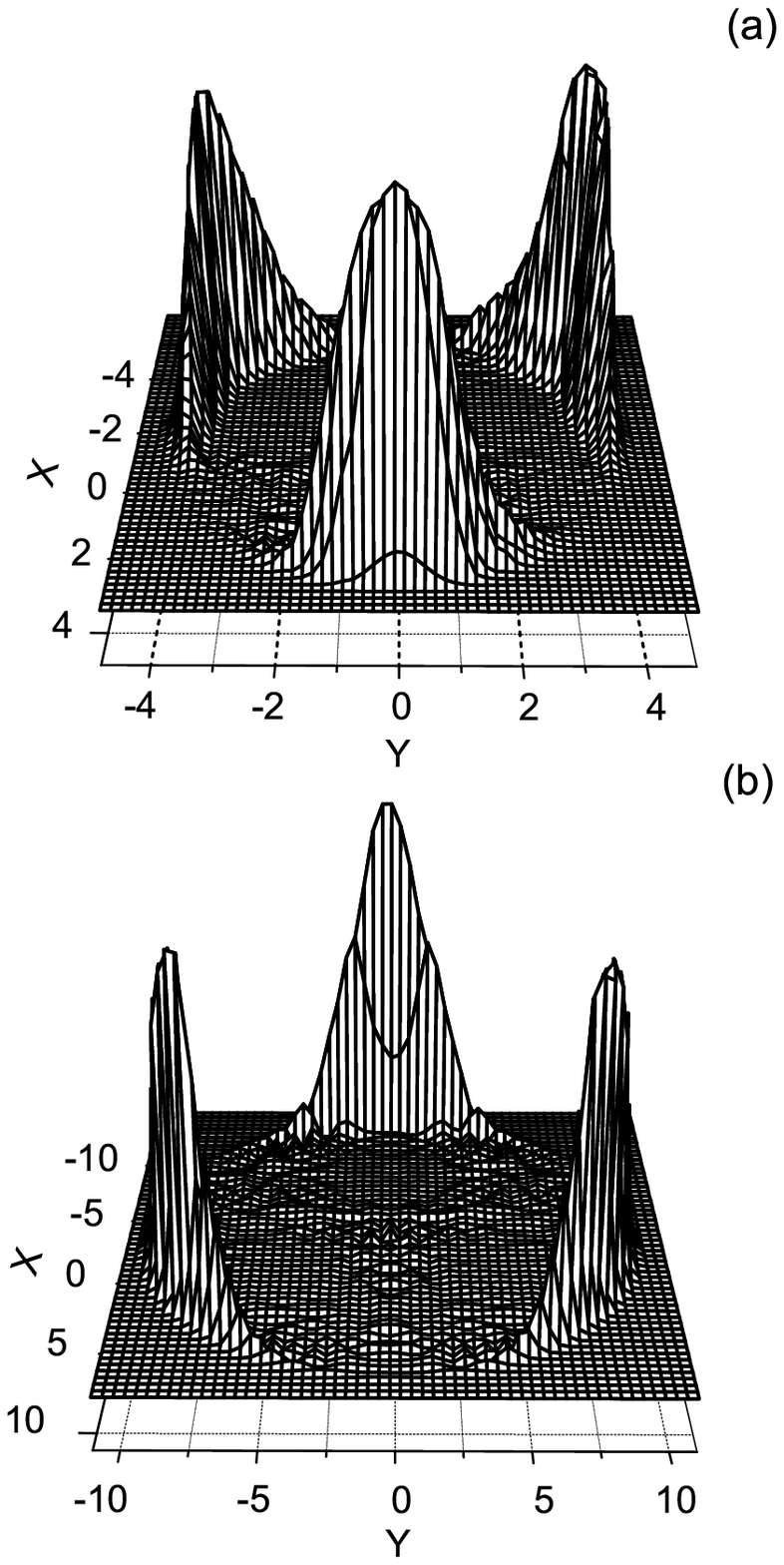}
\caption{Wigner functions in the operational regime above
threshold $E/E_{th}=1.4$ for the parameters:
$\zeta^{'}/\gamma=0.2$, $\xi^{'}/\gamma=0.1$,
$\gamma_{2}=\gamma_{1}=\gamma$. (a) depicts the Wigner function
for the mode $\omega_{2}$, while (b) depicts the Wigner function
for the mode $\omega_{1}$.}\label{Wig12}
\end{figure}

\subsection{Transition through the generation threshold}

We continue the analysis of the sub-section A to include
dissipative effects in quantum distribution of generated mode
$\omega_{1}$ and $\omega_{2}$. For this goal we concentrate on OPO
operated in over transient regime, $t \gg \gamma^{-1}_{1},
\gamma^{-1}_{2}$. The results of the numerical calculations of the
photon-number distributions $P_{1}(n)$ and $P_{2}(n)$ of two modes
(1) and (2) are depicted on Fig.\ref{Pns}. As we see, below
threshold (Fig.\ref{Pns}(a)) these distributions approximately
describe the statistics of spontaneous radiation. It is evident
that the locations of extremum of $P(n)$-function, i.e. the
locations of the most probable values of $n$, may be identified
with the semiclassical stable states (\ref{n1}),(\ref{n2}) in the
limit of small quantum noise level $(\zeta^{'} / \gamma_{1} \ll 1,
\xi^{'} / \gamma_{1} \ll 1 )$. With the increase of these ratios,
due to the multiplicative character \cite{Horsthemke} of the noise
in our system, the locations of these extremum becomes slightly
shifted from the corresponding semiclassical values $n_{1}$ and
$n_{2}$ (\ref{n1}),(\ref{n2}). As shows such comparison, the
Fig.\ref{Pns}(c) exactly demonstrates this situation.

Another remarkable feature of cascaded OPO is formation of
phase-locked states for both subharmonic modes equally separated
by $2\pi /3$ on phase space. As follows from the semiclassical
results (\ref{n1}), (\ref{n2}) there exist three states of equal
intensities, but with different phases. We present below the
results of numerical calculations on phase-locking in quantum
theory in the framework of the Wigner functions. In below
threshold regime solutions of the semiclassical equations for
amplitudes of subharmonics are equal to zero, i.e. both modes are
in the vacuum state, therefore the corresponding Wigner functions
are almost Gaussian below threshold. The results at the threshold
are shown in Fig.~\ref{Wig07}. A clear formation of phase-locked
states in the transition through the generation threshold is seen.
The results above the threshold are shown in Fig.~\ref{Wig12}. In
this operational regime the Wigner functions display three-fold
symmetry in phase-space in accordance with the semiclassical
relations (\ref{n1})-(\ref{FiS}). Note, that quantum interference
pattern has been realized for short interaction time intervals
(see, Fig.\ref{9W} (i)) is disappeared in over transient regime in
the presence of the dissipations.

\section{Conclusion}

In conclusion, we have investigated joint quantum state of
three-photons with arbitrary spectral characteristics generated in
optical superlattises. For this goal two cascaded configurations
have been considered leading to production of spontaneous photon
triplet in cascaded PDC and for generation of high-intensity mode
through cascaded three-photon splitting in optical cavity.
Considering dual-grid structure that involve periodically-poled
crystals we have demonstrated that in cascaded type-II and type-I
configuration triple photons constitutes GHZ entangled
polarization states. We have shown that in short pulsed regime of
cascaded PDC it is possible to control the three-photon joint
spectra by using the method of compensation of the dispersive
effects in non-linear segments by appropriately chosen linear
dispersive segment of superlattice. In the result the production
of heralded joint states of two polarized photons has been
demonstrated in such superlattice by using the conditional method
of detection of auxiliary photon. Considering three-photon
splitting in an optical cavity we have calculated photon-number
distribution and the Wigner function of the mode for short
interaction times. We have shown that in this regime, the Wigner
function displays three-fold symmetry in phase-space
(phase-locking); the three developing arms follow the
semiclassical directions of the phase space with an interference
pattern in the regions between them. We have also investigated
three-photon cascaded down-conversion in cavity in over-transient
regime when dissipative effects are essential. Calculating
third-order correlation function of the photon-number of
sub-harmonic mode we have shown the strong correlation between
three photons registered at the same time-intervals that lead to
three-photon super-bunching in below threshold regime. This effect
decreases if the system moves to the range of generation
threshold. We have also investigated the photon-number
distributions and Wigner functions of sub-harmonics in transition
through the generation threshold.

\begin{acknowledgments}
The research was partly founded by the ISTS (Grant No. A-1517 and
CRDF/NFSAT/SCS (Grant No. ECSP-09-53 A-03). We gratefully
acknowledge these supports.
\end{acknowledgments}

\end{document}